\documentclass[english]{article}
\usepackage[T1]{fontenc}
\usepackage[latin9]{inputenc}
\usepackage{geometry}
\usepackage{color}
\usepackage{graphicx}
\usepackage{amsmath}
\usepackage{amssymb}
\usepackage{babel}


\usepackage{babel}
\begin{document}

\title{New designs of reversible sequential devices}

\author{Anindita Banerjee and Anirban Pathak}

\maketitle
\begin{center}
Jaypee Institute of Information Technology University, Noida, India
\par\end{center}

\begin{abstract}
A clear protocol for synthesis of sequential reversible circuits from any particular gate library  has been
provided. Using that protocol, reversible circuits for SR latch, D latch, JK latch and T latch are designed from
NCT gate library. All the circuits have been optimized with the help of existing local optimization algorithms
(e.g. template matching, moving rule and deletion rule). It has been shown that the present proposals have lower
gate complexities, lower number of garbage bits, lower quantum cost  and lower number of feedback loops compared
to the earlier proposals. For a fair comparison, the optimized sequential circuits have been compared with the
earlier proposals for the same after converting the earlier proposed circuits into equivalent NCT circuits.
Further, we have shown that the advantage in gate count obtained in some of the earlier proposals by
introduction of New gates is an artifact and if it is allowed then every reversible circuit block can be reduced
to a single gate. In this context, some important conceptual issues related to the designing and optimization of
sequential reversible circuits have been addressed. A protocol for minimization of quantum cost of reversible
circuit has also been proposed here.

\end{abstract}

\section{Introduction}

Landauer's principle \cite{Landeur} states that any logically irreversible operation on information, such as the
erasure of a bit or the merging of two computation paths, is always associated with an increase of entropy of
the non-information bearing degrees of freedom of the information processing apparatus or its environment. Each
bit of lost information will lead to the release of at least $kTln2$ amount of heat, where
$k=1.3806505\times10^{-23}m^{2}kg^{2}K^{-1}$$(JoulesKelvin^{-1})$ is Boltzamann's constant and T is the absolute
temperature at which the operation is performed. By 2020 this loss will become a substantial part of energy
dissipation in VLSI circuits, if Moore's law continues to be in effect. This particular problem of VLSI
designing was realized by Feynman and Bennet in 1970s. In 1973 Bennet \cite{Bennet} had shown that energy
dissipation problem of VLSI circuits can be circumvented by using reversible logic. This is so because
reversible computation does not require to erase any bit of information and consequently it does not dissipate
any energy for computation. Reversible computation requires reversible logic circuits and synthesis of
reversible logic circuits differs significantly from its irreversible counter part because of different factors
\cite{Nielsen}. The technological requirement of designing of energy dissipation free VLSI circuits, particular
characteristics of synthesis and testing of reversible circuits and the tremendous advantage of quantum circuits
have motivated scientists and engineers from various background (eg. Physics, Electronics, Computer science,
Mathematics, Material science, Chemistry) to study various aspects of reversible circuits.

Quantum mechanical operations are always reversible and consequently all quantum gates are reversible. A
classical reversible gate can not handle superposition of states (qubit) so it forms a special case of quantum
circuit or a subset of the set of the quantum circuits. But from the construction point of view classical
reversible gates are easy to build \cite{Vos1, Vos2}. A lot of interesting works have already been  reported  in
the field of synthesis  \cite{Kerntopf}-\cite{Mohammadi1}, optimization \cite{Miller, Maslov1}, evaluation
\cite{Mohammadi2} and testing \cite{Vasudevan} of reversible circuits. In a short period the reversible
computation has emerged as a promising technology having applications in low power CMOS \cite{Schrom},
nanotechnology \cite{Merkle}, optical computing \cite{Knill}, optical information processing, DNA computing
\cite{Harlan}, bioinformatics, digital signal processing and quantum computing
 \cite{Nielsen}. From these wide range of potential applications and from the energy dissipation problem of VLSI it is clearly evident  that the reversibility will play dominant role in designing  circuits in future.

 But the designing aspect of reversible sequential circuit is not yet studied rigorously.
This is because of the fact that feedback in a reversible circuit can not be visualized in the usual sense in
which feedback is visualized in a conventional irreversible circuit. This issue was first addressed by Toffoli
\cite{Toffoli1},  where he had shown that the reversible sequential circuits can be constructed provided the
transition function of the circuit block without the feedback loop is unitary. His ideas on the sequential
reversible circuit had further strengthen in his pioneering work on conservative logic \cite{Toffoli2}. Later on
some efforts have been made to construct reversible sequential circuit \cite{Picton}-\cite{Chuang}. All these
efforts are concentrated on the designing of various flip flops because of the fact that the flip flops are the
basic building block of the memory element of a computer and if one wishes to build a reversible classical
computer then these designs will play a crucial role. But several conceptual issues related to designing and
optimization of reversible sequential  circuits are not addressed till now. The earlier works
\cite{Picton}-\cite{Chuang} neither provide any clear protocol for synthesis of reversible sequential circuit
nor they have systematically tried to think beyond the scope of classical irreversible logic. Further  the
quantum cost of the reversible sequential elements have not been calculated so far nor a systematic algorithm
for minimization of quantum cost has been reported. These facts have motivated us to provide a protocol for
synthesis of optimized reversible circuits. We have also provided a protocol  to optimize the quantum cost of a
reversible circuit/gate. We have also addressed the other important conceptual issues related to the designing
of reversible sequential circuits. In the next section we provide background of reversible circuits and  in
section 3 we address conceptual issues related to the feedback and the choice of gate library. In section 4 we
have discussed the earlier approaches and their limitations. In section 5, we have provided a protocol for
synthesis of reversible circuits  we have also used it to  design reversible sequential elements. In section 6,
we have given a protocol for comparing our designs with the earlier proposals. Finally we conclude in section 7.

\section{Background}

A reversible logic circuit comprises of reversible gates. A gate is reversible if it has equal number of inputs
and outputs and the boolean function that maps the input in output is bijective. Consider Fig. 1, where input
vector $I$ is $(x_{1},x_{2},......x_{n})$ and output vector $O$ is $(y_{1},y_{2},......y_{n})$. The gate
(function) is reversible if it satisfies  the condition of one to one and
onto mapping between input and output domains. %

\begin{figure}
      \centering
\includegraphics[scale=0.6]{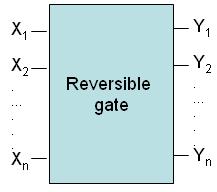}
\caption{Reversible gate}\label{Fig1}
\end{figure}

A garbage bit is the additional output to make a function reversible and it is not used for further
computations. Therefore large number of garbage bits are undesirable in a reversible circuit. As an example, in
Table 1 we have shown an irreversible and a reversible AND gate. It is evident that the Z output gives us the
required output and the other outputs X and Y are garbage.

The quantum cost \cite{Mohammadi2, Maslov2, Barenco, Pallav} of a reversible gate is the number of primitive
quantum gates needed to implement the gate. All  $\left(1\times1\right)$ and $\left(2\times2\right)$ are
considered as quantum primitive gate and the cost all quantum primitive gates are considered. For example we can
construct Toffoli with square root of not gate (V) and CNOT and in that construction the total gate count of
Toffoli is five \cite{Smolin}. Thus the quantum cost of Toffoli is five.
\begin{table}
\centering
\includegraphics[scale=0.7]{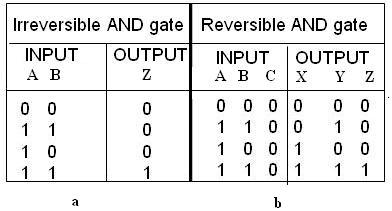}
\caption{And gate; a. Irreversible and b. Reversible }
\end{table}

 Following two methods
have been provided by Mohammadi \cite{Mohammadi3} to find the quantum cost of a non-primitive reversible gate or
a circuit:

1. Implement a circuit/gate using only the quantum primitive  $\left(1\times1\right)$ and
$\left(2\times2\right)$  gates and count them \cite{Maslov2, Barenco, Mohammadi3}.

2. Synthesize the new circuit/gate using the well known gates whose quantum cost is specified and add up their
quantum cost to calculate total quantum cost \cite{Mohammadi3, Islam}.

At this point we would like to mention that quantum cost obtained in these two procedures may be higher than the
actual one unless local optimization algorithm is applied to equivalent circuit obtained in terms of quantum
primitive gates. Further we would like to mention that there is a conceptual difference between optimization
algorithm used for reduction of circuit complexity and the one used for reduction of quantum cost. This is so
because in case of circuit optimization we are restricted to a gate library but to reduce the quantum cost we
can introduce any New gate as long as the gate is $\left(1\times1\right)$ or $\left(2\times2\right)$. Let us
show how the modified local optimization algorithm may help. Consider a Fredkin gate as given in Fig 2a. which
has 3 Toffoli gates. This can further be reduced by template matching to one Toffoli and two CNOT gates as shown
in Fig. 2b. If we substitute the Toffoli gates by quantum primitives we obtain the circuit shown in Fig. 2c.
According to Mohammadi's methods \cite{Mohammadi3} the quantum cost is seven. We now apply the moving rule
\cite{Miller} twice to circuit in Fig 2c (the movements are shown by arrows) to obtain Fig. 2d, in which
 the quantum cost of Fredkin gate is found to be five. Here we would like to draw your attention towards the fact that the moving rule (which was essentially designed to reduce circuit complexity) has not reduced the circuit
complexity but it has reduced the quantum cost.

\begin{figure}
\centering
\includegraphics[scale=0.6]{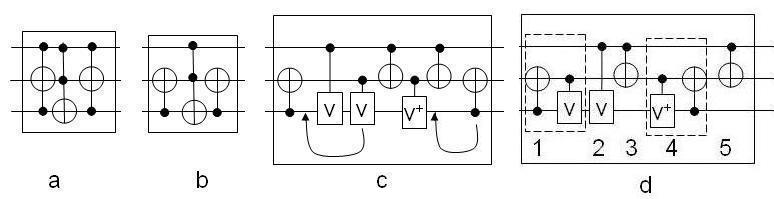}
\caption{Modified local optimization algorithm (moving rule) is applied to
 minimize the  quantum cost of Fredkin gate}\label{Fig 2}
\end{figure}

Thus we have established that local optimization play a very crucial role in reducing the quantum cost, earlier
works \cite{Mohammadi2, Mohammadi3, Islam} have not provided adequate attention towards this fact and
consequently we observe that quantum cost of several gates proposed in earlier works on reversible circuits can
be reduced using local optimization algorithms. Here we would like to note that the recently Maslov
\cite{Maslov1} has used local optimization technique based on templates  to reduce the quantum circuit cost but
he has not provided any systematic protocol for reduction of quantum cost.

 \begin{figure}
     \centering
      \includegraphics[scale=0.95]{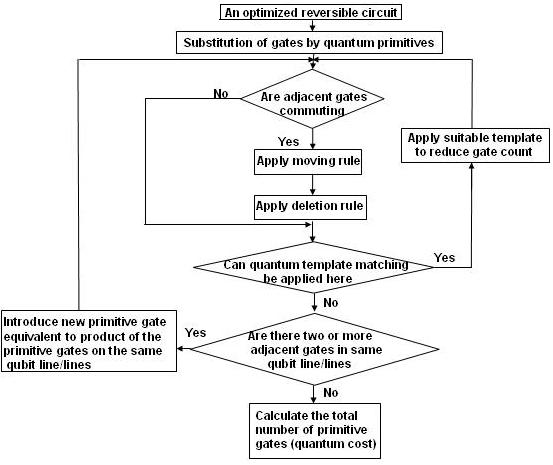}
     \caption{Protocol for optimization of quantum cost}
     \label{Fig 3}
   \end{figure}

We have shown our protocol in Fig. 3 and have used it to find quantum cost of our circuit and also the quantum
cost of different reversible circuits proposed in \cite{Maslov2, Mohammadi3, Islam}.

\section{Conceptual issues related to reversible circuit}

To provide a systematic protocol for designing reversible sequential circuit and to compare the proposed circuit
designs with the existing designs we need to address certain conceptual issues related to reversible circuit
designing. To be precise, conceptual issues related to feedback, choices of gate library and approximate
optimization (local optimization) techniques will be addressed in the following subsections.

\subsection{Feedback in a reversible circuit}

\begin{figure}[h]
\centering \scalebox{0.6}{\includegraphics{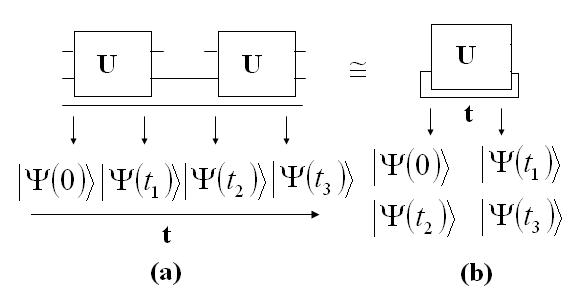}}
 \caption{Circuit a and b are equivalent but the idea
of time axis is not valid in b. Here $\left|\Psi(t)\right\rangle $ is the product state at time t and
$t_{3}>t_{2}>t_{1}>0 $}\label{Fig 4}.
\end{figure}

It is widely believed that feedback is not allowed in a reversible circuit. This is true if we consider feedback
in a similar fashion as it is dealt in classical irreversible logic. The objection against feedback is twofold.
Firstly, merging of two computational paths is not allowed in a reversible circuit and secondly, time axis goes
from left to right in a reversible circuit (as shown in Fig. 4a). Thus if we need to follow the same notion of
time axis in a reversible sequential circuit then feedback will essentially mean a journey in negative time axis
or existence of time machine. This is against the notion of physical reality. But these strong objections
against feedback in reversible circuit can be circumvented by establishing the equivalence between the circuits
in Fig. 4a and Fig. 4b. To be precise, the feedback loop shown in Fig. 4b is only in space not in time.
Therefore, the circuit in Fig. 4b is equivalent to a cascaded circuit in time axis (see Fig. 4a). Thus the usual
notion of time axis is not valid in reversible sequential circuit (i.e. in a circuit having spatial feedback
loop similar to one shown in Fig. 4b). Further, since the circuit in Fig. 4b is equivalent to the cascade shown
in Fig 4a, there is no merging of computational paths and consequently there would not be any loss of energy
provided U is unitary. This conclusion coincides with the Toffoli's idea \cite{Toffoli1} of unitary transition
function. Now if we follow, this notion of feedback, then to establish the reversibility of the design it would
be sufficient to establish the unitarity of U. Here we would also like to note that in this restricted notion of
spatial feedback we can not allow any arbitrary feedback loop. An allowed loop has to be reducible to the
structure shown in Fig. 4b.

\subsection{Latch and Flip flop}

In digital designing a latch is defined as a bistable memory unit which changes its output with the input and is
independent of a clock. The term flip flop is  defined as a sequential device that samples its input and changes
its output only at times determined by clock signals. But this is also called gated latch in some books
\cite{Mano, Sasao}. In these books flip flop are constructed from two gated latches, one of them is the master
latch and other is the slave latch. They have classified flip flops into two categories one is master slave flip
flop. In this flip flop the master latch is disabled when (clock pulse) CP=0 and slave latch is disabled when
CP=1. The other type of flip flop is edge triggered flip flop and this synchronizes the state changes  during a
CP transition.   The structure of master slave type  and edge triggered type are same that is it consists of two
gated latches with a not gate as shown in Fig 7b. This methodology has been followed throughout the existing
designs for  \cite{Rice1, Chuang}. Hence for reversible sequential circuit designing, it is important to mention
which convention is used but hardly any previous works has mentioned this.

\subsection{Gate library: Which gates should be used for the synthesis of the
reversible circuit?}

Whichever synthesis algorithm we follow, it is important to chose a gate library which is universal but the
choice of the gate library (i.e. gates which are the member of that library) is not unique and there does not
exist any single convention. The physical complexity of gates may not be same in two different implementation of
reversible circuits. For example, it may be easy to build an arbitrary gate 'A' in MOSFET technology but it may
not be that easy to implement in optical based technology \cite{Brien, Fiuraek}. A N-qubit reversible gate is
represented by $2^{N}$x $2^{N}$ unitary matrix and product of any arbitrary number of unitary matrices is always
unitary. Consequently, if we put a set of reversible quantum gates in a black box then an unitary matrix will
represent the box and one can technically consider it as a New gate. If we allow such construction of New gates
then any circuit block (of arbitrary size) can be reduced to a single New gate. \footnote{This is true in case
of any quantum circuit block too, provided it does not contain any measurement operation.}

Thus it is straightforward to observe that the use of New gate to reduce the gate count
\cite{Thap4}-\cite{Majid3} is an artifact. To be precise, we would like to mention that to reduce the gate
complexity and garbage bits various papers have reported different New gates. For example,  Hassan Babu
\cite{Hassan} has introduced a 3x3 New gate for full adder circuit, H. Thapliyal  \cite{Thap1, Thap4} has
introduced New gate and 3x3 TKS gate for multiplexer based full adder and multipliers and a 4x4 TSG gate for
carry look-ahead and other adder architectures. In  \cite{Thap5} he has also  introduced New Toffoli gate and
New Fredkin gate for SR latch and JK Gated latch respectively.  Majid Haghparast \cite{Majid1}-\cite{Majid3} has
proposed various gates, which includes New Fault Tolerant gate for fault tolerant reversible logic circuits, HNG
gate for reversible multiplier circuit and MKG gate for full adder. Thus the gate count (gate complexity)
reported in these works are misleading and consequently we need a logical approach to construct an universal
gate library which in turn will help us to compare more than one circuit designs proposed for the same purpose.
\begin{table}
      \centering
\includegraphics[scale=0.6]{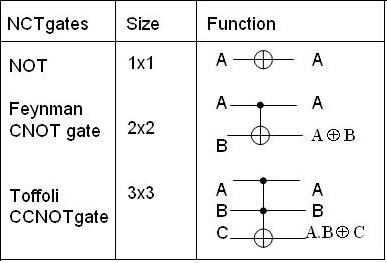}
\caption{Reversible gates from NCT gate library, their size and functions.}
\end{table}

\begin{figure}[h]
\centering
 \scalebox{0.6}{\includegraphics{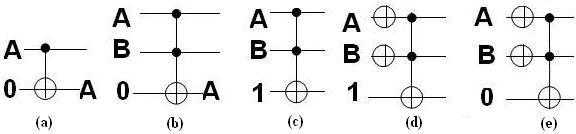}}
\caption{Realization of classical irreversible operations with the help of reversible gates: a) COPY gate, b)
AND gate, c) NAND gate, d) OR gate, e) NOR gate }\label{Fig 5}
\end{figure}
Different universal set of quantum gates have been reported by various groups and among them one of particular
interest was introduced by Aharonov \cite{Aharonov} which contains Hadamard (H) and Toffoli (T) gates. Thus for
all computation in the domain of classical reversible computation it is sufficient to consider Toffoli as the
universal gate. This is easy to visualize through Fig. 5c  where it is shown that the NAND gate can be
constructed using Toffoli. Further we would
 like to note that NOT and CNOT gates can be achieved from (T)
gates. (For NOT gate choose both the first and second control bits as $\left|1\right\rangle $ and for CNOT gate
choose the first control bit as $\left|1\right\rangle $.) Toffoli \cite{Toffoli1} had also shown that the NCT
gate library is universal for the synthesis of reversible Boolean circuits. Actually the inclusion of NOT and
CNOT makes the universal set over complete but this inclusion is justifiable from the perspective of
implementational simplicity  \cite{Vos1, Vos2, Fiuraek} and consistency with earlier works \cite{ Agarwal,
Maslov2}.

Maslov has prescribed a reversible logic synthesis benchmark \cite{Maslov2} in which he has suggested several
gate libraries, which are NCT (NOT, CNOT, Toffoli), NCTSF ( NOT, CNOT, Toffoli, Swap, Fredkin), GT (generalized
n-bit Toffoli) and GT\&GF (generalized Toffoli and generalized n-bit Fredkin). Among these libraries NCT library
is the smallest complete set. Consequently NCT is a good choice of  gate library. Further, these gates can be
experimentally realized using   MOSFET \cite{Vos1, Vos2} and simple optics \cite{Brien, Fiuraek}. Keeping all
these facts in mind, we have chosen NCT gate library. The circuits from NCT gate library as shown in Table 2 is
called NCT circuits. In text we will refer NOT as N, CNOT as C, Toffoli as T. In addition to NCT circuits of
reversible  sequential elements we have also synthesized NCV circuits for the same. The NCV circuits are made
using gates from NCV gate library, which includes  N, C, controlled $V$ and controlled $V^{+}$ gates, is
complete. In \cite{Nielsen} the universality of this library is proved and this library is also used in earlier
work \cite{Maslov1}. Further the NCV circuits are required for determination of quantum cost. Here we would like
to note that the quantum cost of circuit is the number of quantum primitive gates required to construct the
circuit. Since T is not a quantum primitive gate, an NCT circuit can not be used directly to determine the
quantum cost. But T can be constructed using  square root of not gate (V) and CNOT and in that construction the
total gate count of Toffoli is five \cite{Barenco, Smolin}. It is also interesting to note that according to the
Solovay-Kitaev theorem \cite{Kitaev} translation between different universal sets causes only poly-logarithmic
overhead. We have used these facts to compare  our designs of sequential circuits with the existing proposals.
The comparison is done with respect to the NCT gate library, NCV gate library, number of garbage bits (G),
quantum cost (QC) and total cost (TC)\footnote{Sum of garbage bit, quantum cost and gate count (circuit cost)
may be considered as the total cost TC of the circuit.} which is as shown in Table 3, 4 and 5. It should be
noted that reversible logic is used in many areas like optical computing, low power CMOS design, DNA computing
quantum computing etc. The choice of gate library has a significant role in designing implementable feasibility
and resources cost.

\subsection{Why are the optimized designs different?}

Even if we start with the same truth table and same gate library then also different logical paths may lead to
different circuits. After designing a circuit we need to optimize it but an exact optimization technique's time
complexity $(\tau)$ is
\begin{equation}
\tau=\mathcal{O}\left(2^{2n}n^{lm}\right)\label{eq:complexity1}\end{equation} where, $n$ is the number of qubit
line present in the circuit, $m$ is the total number gate present in the circuit and l  is the number of qubit
associated with the largest gate present in the gate library where $l\leq n$. Thus in the present case when the
gate library is (NCT) then $l=3$ and the time complexity of global  optimization algorithm is
\begin{equation} \mathcal{\tau=O}\left(2^{2n}n^{3m}\right)\label{eq:complexity2}\end{equation} This increases
exponentially with $n$ and $m$. In order to avoid this exponential rise in time, certain approximate
optimization (local optimization) algorithms \cite{Miller, Maslov1} have been designed and in practice we use
them. Since the optimization algorithm is an approximate one, it may lead to different circuit designs but the
order of gate complexities have to be the same. This fact is clearly reflected in the Table  3, 4 and 5 given
below.

\section{How to design the circuit}
\subsection{Earlier approaches}

In the previous subsection we have shown that in order to design a reversible sequential circuit we have to
design U as in Fig. 4, as unitary. Now if we know the truth table of U and wish to decompose U in terms of
finite number of logic gates, we can use one of the two existing approaches.
\begin{enumerate}
\item The first approach is the direct substitution method  where one designs reversible circuit \cite{Picton} -
\cite{Thap5} by substituting the irreversible gates  with equivalent reversible gates. To understand this let us
take an example, see Fig 2. in which the classical irreversible gates like OR, NOR, AND, NAND, etc are replaced
by corresponding NCT reversible gates (Fig 2b and 2c) and NCT reversible circuits (Fig 2d and 2e) respectively.
In this approach after designing the equivalent gates one can substitute each irreversible gate of a
conventional circuit by corresponding equivalent reversible gates and obtain the required reversible circuit.
The application of this approach is limited since it requires an existing reversible circuit and it can not go
beyond the limits of classical computation. Picton \cite{Picton} has substituted Fredkin gates in the
traditional SR latch built from NOR gates, Rice  \cite{Rice1, Rice2} has used Toffoli gate in   SR latch and
used Fredkin gate to generate fanout from the clock. Thapliyal  \cite{Thap1} has substituted NAND gate by a New
gate and AND gate by Fredkin gate. In another paper of Thapliyal et al. \cite{Thap5} they have defined another
gate named as New Toffoli gate that replaces NOR gate in traditional SR latch circuit. \item The second approach
is the augmented truth table approach \cite{Chuang} where one starts with irreversible truth table and extend it
to an augmented one and apply transformation based synthesis algorithm to obtain the circuit.
\end{enumerate}
Apart from these two main approaches there are other work  \cite{Thap2, Thap5} where authors have proposed novel
reversible circuits that optimizes the gate count by introducing New gates i.e. they have obtained the
characteristic equation of a particular latch and mapped it to reversible gate (which can be Toffoli or Fredkin)
or proposed a new gate that performs the same logic. Thapliyal \cite{Thap6}  has also designed reversible
latches and flip flops for DNA technology where he has used  Fredkin gates for Fan out, AND and OR operations.

\subsubsection{Limitations of earlier approach }

\begin{enumerate}
\item In the former approach the resource cost is higher which means large number of gates and garbages. This is
so  because each irreversible gate is substituted by a reversible circuit. Also the number of feedback loops are
   same as that in classical irreversible design.
\item The latter approach is not an unique approach, as mentioned in \cite{Chuang}, the designs of latches
depend on the output column (see Table II in \cite{Chuang}), therefore different values assigned to these output
columns will affect the design. Thus it is not clear whether the proposed design is optimal or not.
\end{enumerate}

\subsection{Past works}
 Picton \cite{Picton} was first to propose a reversible circuit for SR latch in 1996. Later on Rice \cite{Rice1} had shown that Picton's design does not have the desired characteristics of SR latch. This observation had motivated Rice to modify Picton's SR latch and   present its Toffoli
version. The design proposed by Rice with Toffoli gates has $\bar{S}$ (Set) and $\bar{R}$ (Reset) as inputs but
while drawing a comparison we consider the primary inputs $S$ and $R$ so we add a NOT gate before S and R and
consider its gate count as 4 as shown in Fig. 6.
\begin{figure}[h]
\centering
\begin{tabular}{cc}
\scalebox{0.6}{\includegraphics{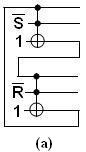}} & \scalebox{0.6}{\includegraphics{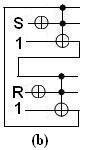}}
\end{tabular}
\caption{Circuits of (a) Rice  SR latch and (b) Rice  SR latch with primary inputs.}\label{Fig 6}
\end{figure}

\begin{figure}[h]
\centering
\begin{tabular}{cc}
\scalebox{0.6}{\includegraphics{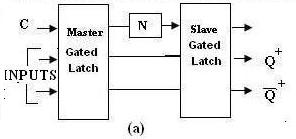}} & \scalebox{0.6}{\includegraphics{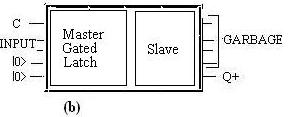}}
\end{tabular}
\caption{Circuits of (a) Conventional master slave flip flop and (b) Reversible master slave flip
flop.}\label{Fig 7}
\end{figure}

 The SR Latch introduced by Thapliyal and Vinod \cite{Thap5} uses a Modified Toffoli gate (see Fig. 9 in  \cite{Thap5}) which was claimed to function as NOR gate for the purpose of designing SR latch  but
we found that it does not function as a NOR gate because the output $R=(A+B)\oplus C$ (where A B and C are the
inputs) so if $C=0$ then $R=(A+B)\oplus0=A+B$. Therefore its implementation gives faulty output. However if we
substitute 0 by 1 at the input i.e.  C=1 then it gives desired output.
 Bijectivity is defined as one to one and on to mapping from the inputs to the outputs
and a reversible gate or circuit should always be bijective. If we operate the traditional flip flop  in Set
condition (i.e. S=1 and R=0)  then the output will be $Q^{+}=1$ and $\overline{Q}^{+}=0$  for following two
cases: (i) when the previous state was in SET condition and (ii) when the previous state was in RESET condition.
Since we obtain same result for two different cases, it always violates bijectivity. This problem may arise in
the circuits designed by the first approach and this fact is reflected in the state table of SR latch reported
by Rice  (see Table V in \cite{Rice1} as it gives same output (0100) for two different inputs (0100) and
(0101)).  Apart from this the circuits of \cite{Picton, Thap1} have fanout problem. Thapliyal and Vinod
\cite{Thap5} presented the transistor implementation of reversible circuit designs which had fanout at various
places of the circuit. Chuang \cite{Chuang} has proposed implementation of sequential elements designed by them
in electron waveguide Y- branch switch technology which also had fanout and fanin at various places.   Apart
from this unstability, is an important concern. All the earlier designs \cite{Picton}-\cite{Thap6} inherited the
unstable condition of the conventional SR Latch which is observed when $S=1$ and $R=1$. Also we would like to
note that all the proposed  reversible flip flops  \cite{Rice1}-\cite{Thap6} (master slave type or edge
triggered type) imitate the same conventional model as given below in Fig. 7. It should be noted that the
purpose of this conventional model of flip flop is to overcome the transparent nature of the gated latch so that
the current output of master is transferred to the slave.
\newpage{}
\begin{figure}[h]
\centering \scalebox{0.9}{\includegraphics{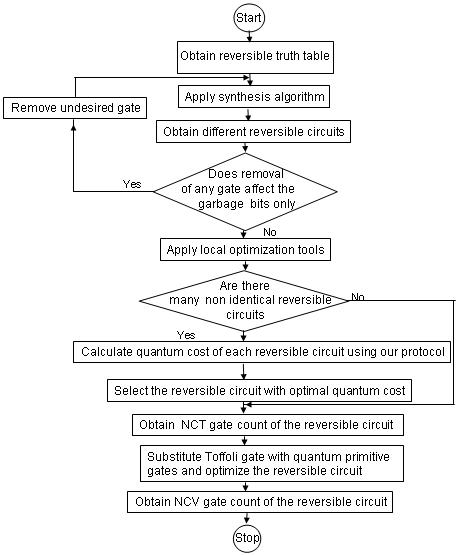}} \caption{Flowchart for the synthesis of reversible
circuits and obtaining  their gate count.} \label{Fig 8}
\end{figure}

\begin{figure}[h]
\centering \scalebox{0.9}{\includegraphics{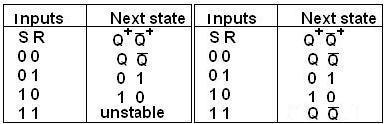}} \caption{Traditional truth table of SR latch and its
new truth table.} \label{Fig 9}
\end{figure}

\begin{figure}[h]
\centering
\begin{tabular}{cc}
\scalebox{0.9}{\includegraphics{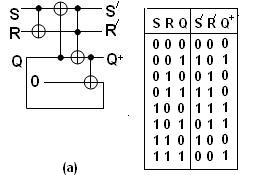}} & \scalebox{0.9}{\includegraphics{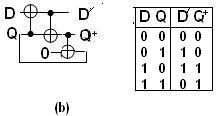}}
\end{tabular}
\caption{Circuit designs of (a) SR latch along with its reversible truth table and (b) D latch along with its
reversible truth table.}\label{Fig 10}
\end{figure}

\begin{figure}[h]
\centering
\begin{tabular}{cc}
\scalebox{0.9}{\includegraphics{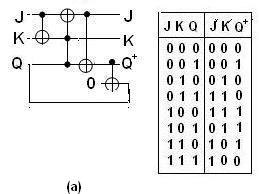}} & \scalebox{0.9}{\includegraphics{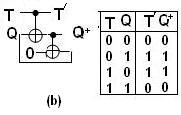}}
\end{tabular}
\caption{Circuit designs of (a) JK latch along with its reversible truth table and (b) T latch along with its
reversible truth table.}\label{Fig 11}
\end{figure}

\begin{figure}[h]
  \scalebox{0.9}{\includegraphics{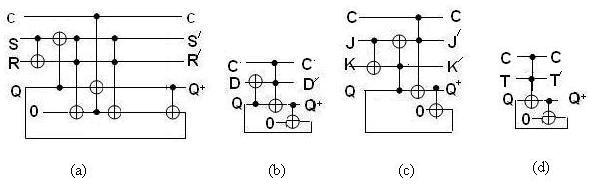}}\\
  \caption{Circuit designs of (a) gated SR Latch, (b) gated D Latch, (c) gated JK Latch and (d) gated T Latch respectively.}\label{Fig 12}
\end{figure}

\begin{figure}[h]
\centering \scalebox{0.9}{\includegraphics{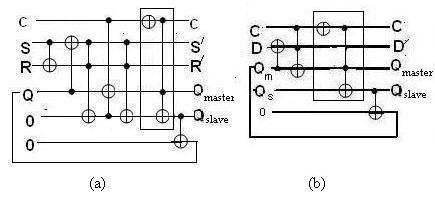}}
\begin{tabular}{c}
 \scalebox{0.9}{\includegraphics{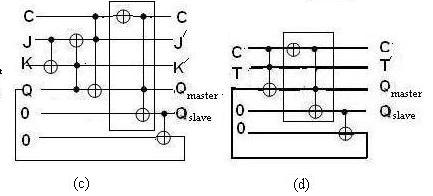}}
\end{tabular}
\caption{Circuit designs of (a) SR Flip flop, (b) D Flip flop, (c) JK Flip flop and (d) T Flip flop
respectively.}\label{Fig 13}
\end{figure}

\begin{table}[h]
\centering \scalebox{0.9}{\includegraphics{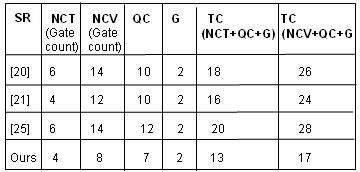}} \caption{Comparison of resources of SR Latch of our
design with existing designs of Picton \cite{Picton}, Rice \cite{Rice1} and  Thapliyal \cite{Thap5}.}
\end{table}

\begin{table}[h]
\centering \scalebox{0.9}{\includegraphics{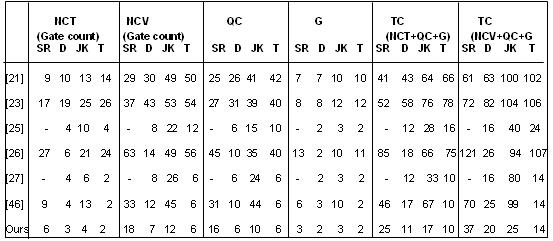}} \caption{Comparison of resources of our designs of
gated latches with  resources of existing designs of Rice \cite{Rice1}, Thapliyal \cite{Thap1, Thap5, Thap6},
Chuang \cite{Chuang} and Banerjee \cite{Anindita}.}
\end{table}

\begin{table}[h]
\centering \scalebox{0.9}{\includegraphics{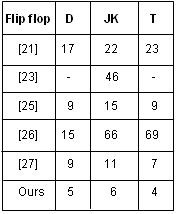}} \caption{Comparison of resources of our designs of
flip flops using  NCT gate library with  resources of existing designs of Rice \cite{Rice1}, Thapliyal
\cite{Thap1, Thap5, Thap6} and Chuang \cite{Chuang}.}
\end{table}

\section{New designs for reversible sequential elements: latch, gated latch and flip flop}

In the earlier works no systematic protocol for synthesis of optimized reversible circuits have been proposed.
Consequently the reversible sequential circuits proposed so far are to some extent adhoc. Here we have proposed
a systematic protocol for the synthesis and  optimization of the reversible sequential circuits. The protocol is
described through the flowchart shown in  Fig. 8. We  have taken a reversible truth table and obtained
different circuits by applying basic algorithm and bi directional algorithm and optimized it. We then calculated
the quantum cost of each circuit and then selected the circuit with optimal quantum cost. We counted NCT gates
and then substituted the T gates with quantum primitive gates and calculated the NCV gate count. In this section
we present some new designs of reversible latches which are constructed using the proposed synthesis protocol.
It was reported by Rice \cite{Rice1, Rice2} that a reversible latch must have at least three inputs which is S,
R and non inverted Q \cite{Rice1}. In our latch we have kept similar inputs and have obtained desired output
$Q^{+}$ while $S^{'}$ and $R^{'}$ are garbage outputs. The traditional truth table of SR latch and its
reversible truth table is given in Fig. 9. It is evident that it does not violate bijectivity. Here as you can
see that the extended reversible truth table is smarter than conventional truth table table because it has no
unstable condition that is  when S and R is 1 then the outputs retain the last state. In other words it works in
a domain which is beyond the reach of irreversible logic.

Also the minimum number of garbage outputs required for reversibility is $\left\lceil log(q)\right\rceil $ where
q is the hamming weight in output patterns (that is the number of times the logical ones is repeated)
\cite{Maslov3}. In our design we have 2 garbage bits, which is minimal. We present some new designs of some
latches along with their truth table in Fig.10 and Fig.11. In designing  latches  we have concentrated  on logic
requirement. For example, in SR latch the requirement would be met by two inputs the output which is feedback to
input that governs the output (output change when inputs are different that is $S\oplus R=1$ and last state and
input $S$ are different otherwise it will hold the last state) so we extended the truth table to logic driven
extended truth table and our inputs S and R are transformed to $S^{'}$ and $R^{'}$. Then we applied
bidirectional algorithm synthesis to obtain a circuit and now our output of $R^{'}$ was $\left(A\oplus
C\right)\oplus\left(C\oplus B\right)=A\oplus B$ so we further optimized it. The algorithm was tried using
different approaches/choice of gate (N, C or T) and different circuits were obtained. Often these circuits have
same NCT gate count even after applying optimization algorithms therefore quantum cost for every circuit is
found and the NCT circuit with optimal quantum cost is selected. Here we would like to note that the number of
loops considerably reduces in above designs. For example the conventional SR latch and proposed SR latch by
Picton \cite{Picton}, Rice \cite{Rice1} and  Thapliyal \cite{Thap5} requires two feed back loops while present
proposal requires one.

We have also presented designs for gated latches which is given in Fig. 12. These have been designed from
respective latches by extending those CNOT gate to Toffoli whose target is $Q^{th}$ bit line.  In Fig. 13 we
have presented the designs for flip flops where  we have used the circuit in box as slave. The slave follows the
master when clock goes low. Consider the SR flip flop in Fig. 13a, here the slave box has a Not gate and a
Toffoli gate which gives the output of flip flop on $4^{th}$  bit line only when the clock goes low. A CNOT gate
at the end copies the output and it is fed back to $3^{rd}$ bit line. We have compared our resources of SR
latch, gated latches and flip flops with the resources used in earlier designs in Table 3, 4 and 5 respectively.

\section{Comparison Protocol}

Since the earlier designs of reversible circuits use different gate libraries. For the purpose of comparison of
circuit complexity of our proposals with the existing proposals we have followed the steps given below:
\begin{enumerate}
\item Equivalent circuit: An equivalent circuit (using NCT gate library) is obtained for each non-NCT gates
using our protocol in Fig. 8 and NCT gate count is obtained,  for example Fredkin gate was used in \cite{Picton,
Rice1} requires 3 NCT-gates, New gate was used in \cite{Hassan} requires 4 NCT-gates, CCCNOT gate was used in
\cite{Rice1, Chuang} requires 3 Toffoli gates, Modified Toffoli gate and Modified Fredkin gate in \cite{Thap5}
requires 3 and 4 NCT gates respectively.

\item Optimization: The equivalent circuits constructed by the above techniques are then optimized with the help
of template matching, moving rule and deletion rule \cite{Miller}.

\item Substitution: Once the optimized circuits equivalent to non-NCT gates are obtained, they are replaced in
the original circuits of Picton, Thapliyal, Rice and Chuang. Thus the essential logic remains the  same.

\item Re optimization: After obtaining the NCT equivalent and logic conserving circuits of earlier proposals,
the optimization techniques (i.e. template matching algorithm, moving rule and deletion rule) are applied once
again on the whole circuit to obtain optimized, NCT equivalent and logic conserving circuit of the earlier
proposals.

\item Cost of resources: Number of NCT gates present in these circuits is counted and this count is considered
as NCT gate count of the circuit. As the choice of NCT as universal gate library is not unique it may be
tempting to see what happens if one uses a different universal gate library. We have chosen NCV as the alternate
universal gate library and calculated the NCV gate count. We have also calculated the quantum cost discussed in
section 2 and finally total cost (TC) of the circuit is obtained by adding the gate count (circuit cost),
quantum cost (QC) and the number of garbage bits (G). While applying moving rule and deletion rule we have often
used following identities:
\begin{enumerate}
\item $V\times V=N$ \item $V\times V^{+}=I$ \item $V^{+}\times V^{+}=N$
\end{enumerate}

\item Comparison: We have compared our resources with the existing resources of  Rice \cite{Rice1}, Thapliyal
\cite{Thap1, Thap5, Thap6}, Chuang \cite{Chuang} and Banerjee \cite{Anindita}  and have found that the present
proposals have lower total cost (TC).
\end{enumerate}
The result of comparison is in Table 3, 4  and 5. It is interesting to note that the advantages of our design
over the earlier proposals (observed for NCT circuits) and all other conclusions remained same.

\section{Conclusions}
In section 3 of the present work, we have addressed the conceptual issues related to the designing and
optimization of reversible circuits in general with a special attention towards the issues related to the
designing of reversible sequential circuits. We have shown that it is very important to define an acceptable
gate library and a good choice for that can be NCT gate library. Further, it has been shown that the advantage
in gate count obtained in some of the earlier proposals by introduction of New gates or unconventional gates
(such as TSG and TKS gates) is an artifact and if it is allowed then every reversible circuit block  can be
reduced to a single gate. We have proposed new designs for SR Latch, D latch, and T Latch with their
corresponding gated latches and flip flops. We have compared our proposals with the existing proposals
\cite{Picton}-\cite{Chuang} and calculated the total cost (TC) of the circuit and found that our circuits use
least resources. We have shown that in appropriate reversible designs one can go beyond the domain of classical
irreversible logic. For example, we have shown that the unstable condition of SR latch is not present here and
minimum number of feedback loops required in the domain of irreversible logic can be reduced in the domain of
reversible logic. To be precise, we have observed that conventional JK flip flop requires 8 feedback loops
whereas reversible Master slave flip flop proposed in \cite{Chuang} requires 2 and our design requires 1. It is
straightforward to realize that number of feedback loop is minimal. We have used NCT gate library and there
exist several proposals \cite{Vos1, Vos2, Brien, Fiuraek} for realization of CNOT and CCNOT gates using optical
computing and CMOS based technology. Thus experimentalists can easily implement our circuits. Consequently, we
can easily build classical reversible memory element. But the implementation of the present work is not limited
to classical domain, this is because of the fact that we can also implement the proposed circuits in quantum
domain with the help of optical implementation \cite{Brien, Fiuraek}.  If one aims to provide optimized
reversible circuits for all the useful components of a classical computer  then this work along with the
proposal of \cite{Vos1, Vos2, Brien, Fiuraek} will help him to provide a complete design for a classical
reversible computer. Since it will be free from the problem of decoherence and scalability it seems more
practical and easy to built than a real scalable quantum computer.

\end{document}